\documentclass{ieeeaccess}
\usepackage{cite}
\usepackage{amsmath,amssymb,amsfonts}
\usepackage{algorithmic}
\usepackage{graphicx}
\usepackage{textcomp}
\usepackage{amssymb}
\usepackage{url}
\usepackage{cite}
\usepackage{algorithm}
\usepackage{algorithmic}
\usepackage{soul}
\usepackage{subfigure}
\usepackage{amsmath}
\usepackage{relsize}
\usepackage{amsmath}

\usepackage[table,xcdraw]{xcolor}
\usepackage{multirow}
\usepackage{csquotes}
\usepackage{subfig}
\usepackage{cleveref}
\usepackage{etoolbox}

\def\BibTeX{{\rm B\kern-.05em{\sc i\kern-.025em b}\kern-.08em
    T\kern-.1667em\lower.7ex\hbox{E}\kern-.125emX}}
\begin{document}
\history{Date of publication xxxx 00, 0000, date of current version xxxx 00, 0000.}
\doi{10.1109/ACCESS.2017.DOI}

\title{A Spatial-Temporal Correlation Approach for Data Reduction in Cluster-Based Sensor Networks}
\author{\uppercase{Gaby Bou Tayeh}\authorrefmark{1},
\uppercase{Abdallah Makhoul\authorrefmark{1},  Charith Perera.\authorrefmark{2}, and  Jacques Demerjian\authorrefmark{3}}}
\address[1]{Femto-St Institute, UMR 6174 CNRS, Universit\'e de Bourgogne Franche-Comt\'e, France}
\address[3]{LARIFA-EDST, Faculty of Sciences, Lebanese University, Fanar, Lebanon}
\address[2]{Cardiff University, Cardiff , United Kingdom}
\tfootnote{This work is partially funded by the EIPHI Graduate School (contract "ANR-17-EURE-0002"), the France-Suisse Interreg RESponSE project,  the Lebanese University Research Program (Number: 4/6132) and EPSRC PETRAS 2 (EP/S035362/1).}

\markboth
{Author \headeretal: Preparation of Papers for IEEE TRANSACTIONS and JOURNALS}
{Author \headeretal: Preparation of Papers for IEEE TRANSACTIONS and JOURNALS}

\corresp{Corresponding author: Charith Perera (pererac@cardiff.ac.uk)}

\begin{abstract}

In a resource-constrained Wireless Sensor Networks (WSNs), the optimization of the sampling and the transmission rates of each individual node is a crucial issue. A high volume of redundant data transmitted through the network will result in collisions, data loss, and energy dissipation.
This paper proposes a novel data reduction scheme, that exploits the spatial-temporal correlation among sensor data in order to determine the optimal sampling strategy for the deployed sensor nodes. This strategy reduces the overall sampling/transmission rates while preserving the quality of the data. 
Moreover, a back-end reconstruction algorithm is deployed on the workstation (Sink). This algorithm can reproduce the data that have not been sampled by finding the spatial and temporal correlation among the reported data set, and filling the \enquote{non-sampled}
parts with predictions. We have used real sensor data of a network that was deployed at the Grand-St-Bernard pass located between Switzerland and Italy. We tested our approach using the previously mentioned data-set and compared it to a recent adaptive sampling based data reduction approach. The obtained results show that our proposed method consumes up to $60\%$ less energy and can handle non-stationary data more effectively.

\end{abstract}

\begin{keywords}
Wireless sensor networks , Data reconstruction , spatial-temporal correlation, Data reduction
\end{keywords}

\titlepgskip=-15pt

\maketitle

\section{Introduction}
\label{sec:introduction}
\PARstart{T}{he} momentum and growth of large-scale sensor networks have been increasing over the recent years. The rising popularity of such networks is due to the fact that they can be used in numerous and diverse event monitoring applications including traffic, air and water quality, e-health, environmental monitoring (wildlife, forest fires, storms, etc.), and many other applications. Such networks are expected to operate autonomously and for a long period of time. However, in a large scale sensor networks, the high volume of redundant data being communicated through the network increases collision, causes data loss, and most importantly it costs sensor nodes a large amount of scarce energy resources.
Therefore, due to severe energy, computational and bandwidth constraints, a sound body of literature has centered on optimizing the efficiency of both the sensing and transmitting 
activities in order to maximize the lifetime of the network.

One of the most commonly used approaches to tackle this problem is the sampling rate adaptation \cite{eSampling,Asampling2,TIIMakhoul,AdhocMak}. A sampling rate is a rate at which a new sample is taken from a continuous signal provided by the sensor board. This rate can be adapted according to the input acquired from the monitoring area. If no significant change is noticed for a certain period of time, the sampling rate could be reduced for the upcoming period, and in contrast, if an event is detected, the sampling rate is increased. This sampling rate adaptation is based on event detection \cite{eSampling,ASTR}. Another sampling rate adaption technique takes into consideration the temporal and spatial correlation among the reported data \cite{Asampling2}, and limits the sampling rate of the sensors that show high correlation with other neighboring ones, and maximizes the sampling rate of those showing a little or no correlation at all. Both approaches aim to reduce the amount of redundant data being transferred through the network.

Other data reduction approaches focus solely on reducing the number of transmissions while maintaining a fixed sampling rate \cite{HLMS,LMS,OSSLMS,TR,CP1}. The most popular of them all is the dual prediction scheme. A prediction model capable of forecasting future values is trained and shared between the source and the destination, thus enabling the source sensor node to transmit only the samples that do not match the predicted value. Some approaches also combine both adaptive sampling and transmission reduction into a single mechanism \cite{DPCAS} aiming to minimize further energy consumption.

In this paper, we propose a spatial-temporal Correlation based Approach for Sampling and Transmission rate Adaptation (STCSTA) in cluster-based sensor networks. The sensor nodes do not need to run any algorithm. The cluster head is responsible for collecting data from its member sensor nodes, computing a correlation function in order to measure the correlation degree among these nodes. Finally, the sensors that show high correlation will be asked to reduce their sampling rate and the ones showing low correlation will be asked to increase it. Moreover, in order to ensure the integrity of the data, a reconstruction algorithm deployed on the Sink station. The latter is used to reconstruct the \enquote{non-sampled} measurements by exploiting the temporal and spatial correlation among the reported data. We compare our approach to a Data Prediction with Cubic Adaptive Sampling (DPCAS) and to an exponential Double Smoothing-based Adaptive Sampling (EDSAS) using real sensor data. The latter and the former combines both adaptive sampling and transmission reduction into a single mechanism, allowing us to compare the efficiency of our proposal with two very effective approaches in terms of reducing radio communication.

The rest of the paper is organized as follows: In section~\ref{RW}, the work related to energy efficient data reduction in a wireless sensor network is presented. In section~\ref{SandEmodel} the system model is briefly explained and the energy model to calculate the energy consumption is illustrated. A detailed explanation of the proposed approach is provided in section \ref{STCSTA}, while experimental results are discussed in Section~\ref{ER}. This paper ends with a conclusion section, in which the contribution is summarized and intended future work is outlined.

\section{Related Work}
\label{RW}

Resource management in sensor networks is a widely discussed topic among researchers. Subsequently, there have been numerous studies regarding this topic. In this section, we present and discuss the different approaches used to tackle this issue.

Compression~\cite{Compression1,compression3,JA1,JA2} and aggregation~\cite{Agg1, Agg1Mak, Agg2Mak} are two techniques aiming to reduce the amount of data routed through the network \cite{CP2}.
The former focus on compressing the data before transmission to the upper node in the network hierarchy and the latter filters and clean the data by removing redundant information before routing these data to the Sink station. Several data compression and aggregation techniques have been proposed in the literature. The authors in~\cite{Compression1} proposed a compression technique for sensor networks organized in a cluster topology. The approach called Cluster-Based Compressive Sensing Data Collection (CCS) compresses data on the cluster head level by generating Compressive Sensing (CS) measurements based on block diagonal matrices created from the raw data received from neighboring sensors. Moreover, the compressed CS measurements are finally reconstructed at the base station (Sink). 
In~\cite{compression3} the authors proposed a compression scheme called Compressive Data Collection (CDC) for Wireless Sensor Networks, it exploits the spatial-temporal correlation among sensory data to perform compression. The scheme consists of two layers, the opportunistic routing with compression and the nonuniform random projection based estimation for reconstruction.
The authors in~\cite{Agg2} proposed a data aggregation technique called the Prefix-Frequency Filtering (PFF). This approach mainly consists of two aggregation layers, the first one is on the sensor level, and the second one is on the cluster head or the aggregator. On both layers, redundant measurements are filtered using the Jaccard similarity that measures the correlation among collected measurements.
In~\cite{Agg1} a Dynamical Message List Based Data Aggregation (DMLDA) technique is presented, it is based on a special data structure called dynamical list. The latter stores the history of received measurements, that are then used to filter any duplicates.

One of the most energy consuming activities in WSN beside transmission and processing is sampling, therefore several studies have been conducted on how to reduce the amount of sampled data through a technique known as \enquote{adaptive sampling}, where a sensor can adapt its sampling rate according to the change in the input environment. The authors in~\cite{eSampling} proposed and event-sensitive adaptive sampling and low-cost monitoring (e-Sampling) scheme, where each sensor has short and recurrent bursts of high sampling rate in addition to a low sampling rate. Depending on the analysis of the frequency content of the signal, each sensor can autonomously switch between the two sampling speed. The authors in~\cite{Asampling2} presents a decentralized temporal correlation based adaptive sampling approach, enabling each sensor to decide its own sampling rate while controlling the size of the sampling interval by limiting the interval size to an automatically calculated \enquote{MaximumSkipSamplesLimit (MSSL)} value. 

The overwhelming majority of studies agree on the fact that radio transmission is the most consuming activity in WSN~\cite{EnergyModel,En2,En3}. Accordingly, numerous studies focused on developing techniques to limit the number of radio transmissions. Most of these techniques are based on the concept of data prediction. The idea is to build on the Sink a prediction model using previously collected readings, that is capable of forecasting future measurements. Enabling the sensor node to transmit a reading only when the prediction does not respect the error tolerance predefined the user.
The authors in~\cite{HLMS} proposed a Hierarchical Least Mean Squares (HLMS) adaptive filter as a prediction model, which is one of the many adaptive filter based approaches \cite{RLS,LMS,OSSLMS}.
This filter consists of multiple layers of regular Least Mean Square (LMS) filters, each layer takes feedback from the previous layer in the hierarchy aiming to minimize the prediction error of the model. Another technique called Derivative Based Prediction (DBP) was introduced in \cite{DBP}, it is less complex than the adaptive-filter based methods. The prediction model is simply a straight line that interpolates a fixed window of data of size $m$ using the first and last $l$ values in the window. In \cite{DPCAS} the authors proposed an approach that combines an adaptive sampling method that is based on the TCP CUBIC congestion protocol, with a transmission reduction method that is based on an exponential predictive mode. The complete data set including the \enquote{non-sampled} and \enquote{non transmitted} measurements are then reproduced on the sink by interpolating the received measurements.
This approach was inspired by both \cite{ASTCP}, and \cite{EDSAS}. The latter
uses an exponential Double Smoothing-based Adaptive Sampling (EDSAS) technique, that adapts the sampling rate of a sensor based on the accuracy of a prediction model. As long as this model is producing good predictions the sampling rate is kept low. It is increased, however, when the predictions surpass a predefined error threshold. The former operates in similar fashion, more specifically it adopted the TCP congestion control to adapt the sampling rate of the sensor node. Thus the approach is called Adaptive sampling TCP (ASTCP).

Both compression and aggregation are effective in term of reducing the data load in the network, however, their performance is limited and cannot reach the efficiency of techniques such as adaptive sampling and transmission reduction. Therefore, compression and aggregation are considered to be as a complementary layer that can be added to adaptive sampling and transmission reduction to further increase their efficiency.
Despite being very effective in reducing the amount of sampled and transmitted data, adaptive sampling and transmission reduction techniques can still consume a substantial amount of energy. This is proportionally related to the complexity of the algorithms that are required to be implemented on the sensor level. The CPU running complex algorithms can consume more energy than the sampling activity~\cite{EnergyModel}, which renders the adaptive sampling technique obsolete in case the implemented algorithm requires a large number of CPU cycles.

In order to schedule the sampling intervals of sensor nodes and reduce energy transmission, some approaches rely on the spatial-temporal correlation between sensor nodes deployed in the monitoring area~\cite{ST1, ST2, ST3, ST4, ST5}. 
The Authors in \cite{ST1} proposed an Efficient Data Collection Aware of spatial-temporal Correlation (EAST). In the latter, the sink subdivides the event area into spatially correlated cells of the same size, then, in each cell, the node having the highest residual energy is elected as a representative node. Only the latter transmits data to the sink while also applying a temporal correlation suppression method on its collected data. Finally, at each time instance, the representative node is re-elected according to the same previous rule. The main drawback of this approach is the size of the cell representing an area of spatially correlated nodes is static, and it is not calculated according to the real level of correlation. Moreover, the representative node is chosen according to residual energy rather than its correlation with other nodes in the cell. Therefore, the term \enquote{representative} is not necessarily true.

In \cite{ST2} the authors proposed a sleeping schedule algorithm that aims to minimize the total spatial-temporal coverage redundancy among neighboring nodes while maximizing coverage. Each sensor node compares itself with neighboring ones using a weight criteria and it locally optimizes its scheduling according to its coverage redundancy.
This method requires constant message exchange between sensor nodes in order to keep track of the changing weight of each one of them, which can produce overhead.

The authors in \cite{ST3} proposed a spatial-temporal correlation model that aims to extend
the network lifetime by scheduling a sleeping period for sensors showing high similarities with other ones belonging to the same cluster. The similarity is measured by computing the Euclidean Distance,
Cosine Similarity and Pearson Product-Moment Coefficient (PPMC). If the result of one of the three indicates a high similarity, the sensor node is set to sleep for half of the period time (1 period = N samples). The first problem with such an approach is if a sensor X shows a similarity with a sensor Y, the opposite is also true (sensor Y will show similarity with sensor X), therefore, according to this approach, both sensors will be set to sleep. By doing so correlated sensors will miss simultaneously instead of compensating for one another by keeping one of them awake.
The second problem is that the sleeping duration is static instead of being computed in a dynamic way according to the level of correlation.

Motivate by the problems related to the aforementioned approaches, we present in this paper a spatial-temporal Correlation approach for Sampling and Transmission rate Adaptation (STCSTA) in cluster-based sensor networks.
Our approach does not require any algorithm to be implemented on the sensor level, the only task performed by sensors are uniquely sampling and transmission.
All the work is done on the Cluster-Head (CH) level, where at the end of each round (duration predefined by the user), the CH runs an algorithm that finds the spatial correlation among the data reported by the sensors belonging to the same cluster. Then, it transmits to one of them its new sampling rate for the next round according to its level of correlation with other neighboring sensors in the cluster. 
The sampling rate scheduling respects a strict protocol that keeps the sampling rate of the sensors showing high correlation with a large number of nodes at an optimal maximum level. Moreover, the protocol prevents highly correlated sensors from missing simultaneously, allowing one to compensate for another. in addition to sampling rate scheduling, and in order to ensure the integrity of the data, a reconstruction algorithm is deployed on the Sink. This algorithm can identify the time stamps where data has not been received due to a reduction in the sampling rate of a specific sensor, and reconstruct them using the spatial-temporal relations among the collection of data reported by the sensors.

\section{System and Energy Model}
\label{SandEmodel}

\subsection{System model}
\begin{figure}[]
    \caption{Illustrative example of the network architecture}
    \centering
    \includegraphics[width=\linewidth]{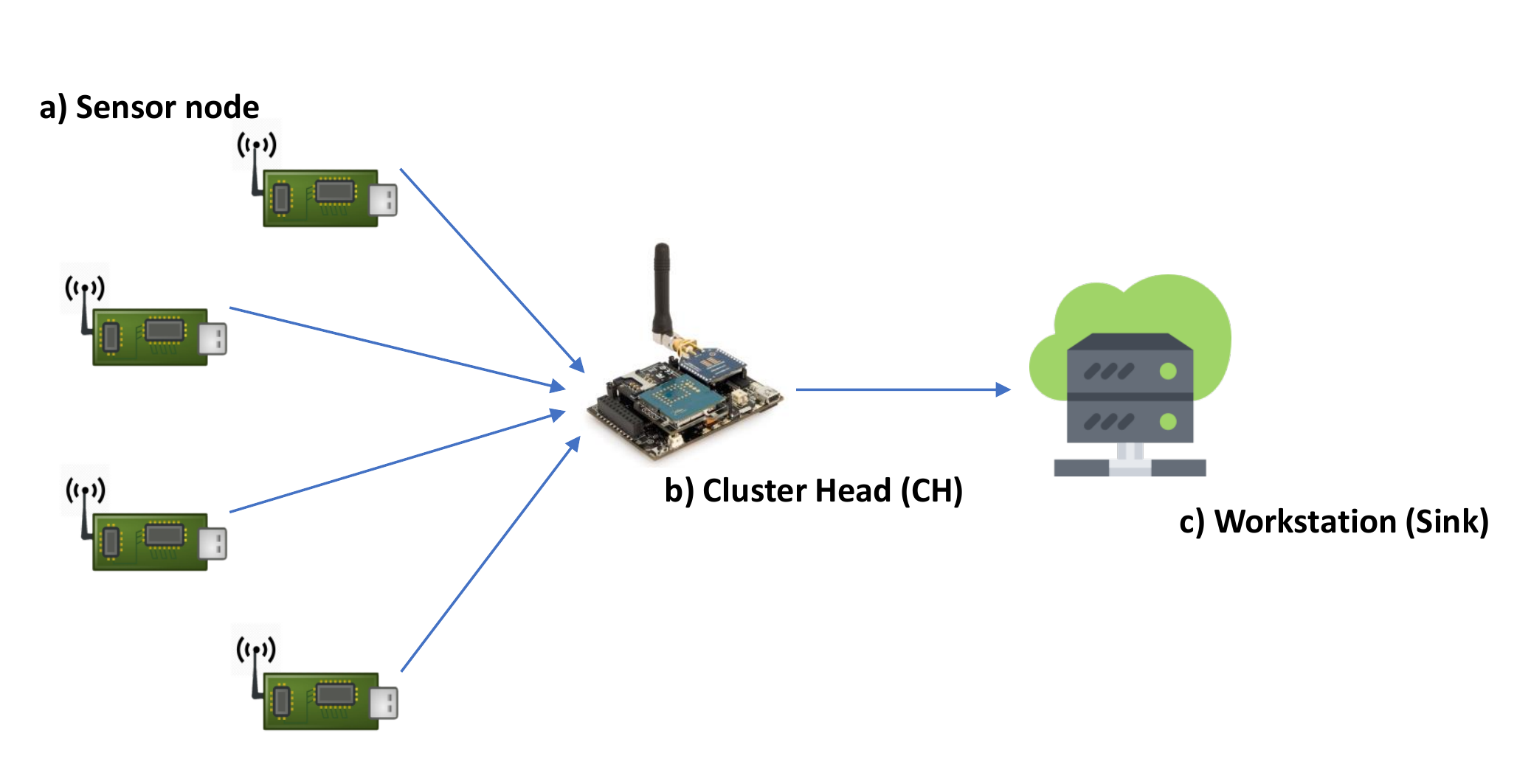}
    \label{fig:Network}
\end{figure}

We consider a set S of N sensor nodes and C cluster heads deployed over a specific monitoring area at locations LS=\{$ls_1,ls_2,...,ls_N$\} and LC=\{$lc_1,lc_2,...,lc_C$\} respectively, where a sensor $S_i$ is located at the location $ls_i$ and a cluster-head $C_j$ is located at the location $lc_j$, and the Sink S is placed in a distant location at a position $l_0$.
Sensor nodes are grouped into clusters, where each one of them belongs to one cluster only.
The cluster heads are considered to be more powerful than sensor nodes in term of processing capabilities and they have been allocated larger energy resources.
Figure~\ref{fig:Network} illustrates an example of the described network architecture for one cluster.

The network is periodic and operates in rounds, where each round R is exactly P seconds, and it is subdivided into m time slots, where at each time slot a sensor samples one measurement. Therefore, the maximum sampling rate ($SR_{max}$) is considered to be P/m samples per round. During the very first round, each sensor node collects data using the maximum sampling rate $SR_{max}$ and transmits the readings to the CH after each acquisition. On the CH level, when the latter receives a measurement from any sensor $S_i$ it stores the values in its memory and routes it directly to the Sink. At the end of the first round, the CH would have stored in his memory the following matrix $M$.
where $n$ is equal to the current sampling rate ($SR_{max}$) in this case, and N is the number of sensors in the cluster.
\begin{center}
$M
    =
    \begin{bmatrix}
    \centering
        x_{1}^{1} & x_{1}^{2} & x_{1}^{3} & \dots  & x_{1}^{n} \\
        x_{2}^{1} & x_{2}^{2} & x_{2}^{3} & \dots  & x_{2}^{n} \\
        \vdots & \vdots & \vdots & \ddots & \vdots \\
        x_{N}^{1} & x_{N}^{2} & x_{N}^{3} & \dots  & x_{N}^{n}
    \end{bmatrix}
    $
\end{center}
The CH than proceeds to computing the correlation between each pair of sensors (The number of possible pairs is $\frac{N(N-1)}{2}$). Using the obtained correlation results the CH calculates than transmit to each sensor node its new SR. A detailed explanation of how the correlation is calculated and how the new SR is determined is provided in section \ref{STCSTA}. 
For the next round, each sensor samples data according to its new sampling rate provided by the CH. 
For Instance, if the latter demands a specific sensor to reduce its sampling rate by 40$\%$, and supposing that $SR_{max}$ is equal to 50 measures/round, the sensor is supposed to sample 30 measurements instead. If each period is 10 minutes long (600s), instead of sampling a measurement every 12 seconds (600/50), the sensor would sample a measurement every 20 sec (600/30). 
Moreover, Knowing the duration of each period, the maximum sampling rate and the time stamp when each measurement was received, both the Sink and the CH are capable of identifying the non-sampled data, which will be replaced by "Nan" (see matrix M$^\prime$) in order to reconstruct them later at the Sink station and in order to make the computation of the correlation among sensor nodes easier for the CH as explained in section~\ref{corr}. Therefore, the stored matrix that is used to compute the correlation will actually be as shown below, where n is equal to the maximum number of samples per round ($SR_{max}$):

\begin{center}
$M^\prime
=
\begin{bmatrix}
\centering
    x_{1}^{1} & x_{1}^{2} & x_{1}^{3} & \dots & x_{1}^{50} &  Nan  & x_{1}^{n} \\
    x_{2}^{1} & x_{2}^{2} & x_{2}^{3} & \dots & Nan & Nan & x_{2}^{n} \\
    \vdots & \vdots & \vdots & Nan & \vdots & \vdots & \vdots \\ 
    x_{N}^{1} & x_{N}^{2} & x_{N}^{3} & \dots & x_{1}^{50} & Nan & x_{N}^{n}
\end{bmatrix}
$
\end{center}

\subsection{Energy model}

In order to compute the energy consumption of a sensor node \cite{CP3, CP4}, it is necessary to take into consideration the energy consumed by every single operation performed by the node. Generally, the consumed energy relates to four main tasks, namely, sampling, logging, processing, and radio transmission. Therefore, the energy consumption model can be defined as:

\begin{equation}
\label{eq:energyEq}
E_{node}=E_{sampling} + E_{logging} + E_{processing} + E_{radio}
\end{equation}
Where $E_{sampling}$ is the energy required for sampling one value, $E_{logging}$ is the required energy to log data in the memory, $E_{proccessing}$ is the required energy to run and algorithm consenting of $N$ CPU cycles, and $E_{radio}$ is the energy required to transmit a $b$ bits packet for a distance $d$.
In this article we use the energy model discussed in \cite{EnergyModel} to calculate the overall energy consumption of each sensor node.

\section{The Proposed Approach (STCSTA)}
\label{STCSTA}
In this section, we will explain in detail, how the correlation between sensor nodes and the new sampling rates of each sensor are calculated.

\subsection{Computing correlation and sampling rate allocation}
\label{corr}

\textbf{Algorithm 1 - line(2-14)} : After a round is completed, each sensor node would have transmitted to the cluster head a different number of measurements since the sampling rate of each one of them can be different. Nevertheless, as mentioned earlier the CH identifies the non sampled data and fill their corresponding place in the vector by a Nan value, therefore all the vectors will have the same size $n$. However, the correlation between two vectors containing Nan values cannot be computed. Therefore, each and every Nan value is replaced by the value of the first \enquote{non-Nan} value that comes before it in the same vector.
For instance, in the \enquote{M$^\prime$} matrix, $x_{1}^{51}$ is Nan it will be set equal to the same value as $x_{1}^{50}$, and $x_{2}^{50}$ and $x_{2}^{51}$ are set equal to the same value as $x_{2}^{49}$, and so on. 

\textbf{Algorithm 1 - line(17-22)} : Afterward, the linear dependency of each pair of vectors ($v_i$,$v_j$) $\in$ M$^\prime$ is calculated using the Pearson correlation coefficient. The latter is known as the best method of measuring the association between variables of interest because it is based on the method of covariance.  It gives information about the magnitude of the association, or correlation, as well as the direction of the relationship.
The Pearson correlation coefficient is described in the equation~\ref{eq:pc} below, where $\mu$ and $\sigma$ are the mean and standard deviations.

\begin{equation}
    \label{eq:pc}
    \rho(v_i,v_j)= \frac{1}{n-1}\times\sum_{k=1}^{n}(\frac{\overline{v_{ik}-\mu_{v_i}}}{\sigma_{v_i}})(\frac{\overline{v_{ik}-\mu_{v_j}}}{\sigma_{v_j}})
\end{equation}

\begin{figure}[]
    \centering
	\includegraphics[width=\linewidth]{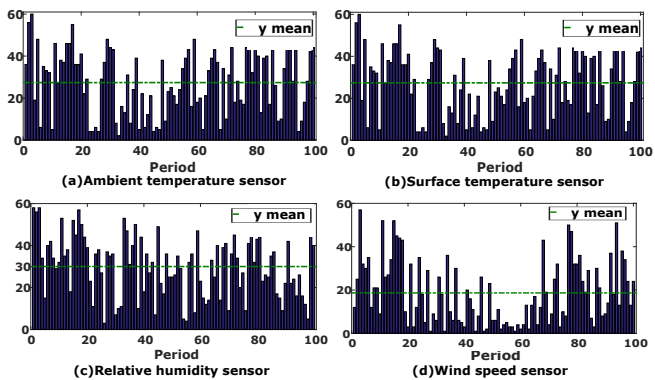}
	\caption{Figure showing the number of moderately $\&$ highly correlated sensors (pearson correlation coefficient $\geq$ 0.5) during each one of the first 100 Periods}
	\label{fig:corr}
\end{figure}

The justification behind using the Pearson correlation can is illustrated in Figure~\ref{fig:corr}. We have used a data set of 92 sensors to generate 4 graphs that show the number of sensors that are moderately $\&$ highly correlated with 4 randomly chosen sensors during each period and for the first 100 periods. For instance, in Figure~\ref{fig:corr}(a) we notice that this randomly chosen ambient temperature sensor correlates with a large number of sensors during each period. On average it correlates with 27 sensors as the mean values shows.
Same for Figure~\ref{fig:corr}(b) and (c) on average these sensors correlate with approximately 30 other sensors that are in the same cluster. However, The mean value in Figure~\ref{fig:corr}(d) is significantly lower (mean=19), in section~\ref{QR} we will see how this will reflect on the results.

Heterogeneous environmental data beside other types of data such as medical data (vital signs), movement tracking data (speed, acceleration, location) and etc, are usually highly and/or moderately correlated.
This correlation thus can be used in order to reduce the number of transmitted measurements by deriving values from other observed ones. This is indeed the motivation behind using correlation to adapt the sampling rate of the sensors.

\textbf{Algorithm 1 - line(23-28)} :After computing the correlation value of each sensor $i$ with all the other sensors belonging to the same cluster, the CH looks for the sensor $j$ that it correlates the most with as shown in table~\ref{table:corr1}. 

\textbf{Algorithm 1 - line(29-38)} :Afterward, the CH counts the number of occurrences of each sensor $j$ in the second column of the table and stores them in a list according to their ascending order. 

\begin{table}[!htb]
\centering
\begin{tabular}{|c|c|c|}
\hline
Sensor i & \begin{tabular}[c]{@{}c@{}}Sensor j \\ (has the max \\ correlation \\ with Sensor i)\end{tabular} & Correlation degree \\ \hline
1        & 54                                                                                                & 0.91        \\ \hline
2        & 7                                                                                                & 0.87        \\ \hline
3        & 5                                                                                                 & 0.70         \\ \hline
4        & 2                                                                                                & 0.96        \\ \hline
5        & 6                                                                                                 & 0.75        \\ \hline
...      & ...                                                                                               & ..          \\ \hline
n        & 32                                                                                                & 0.88        \\ \hline
\end{tabular}
\caption{The correlation table}
\label{table:corr1}
\end{table}

\textbf{Algorithm 1 - line(39-49)} :Starting from the first sensor $j$ in the ordered list, the CH looks in table~\ref{table:corr1} for the sensor $j$ in the first column and extract the value of its max correlation from the third column. Then the CH notifies $j$ that its sampling rate must be reduced proportionally to the correlation value. For instance, if sensor $5$ was first in the ordered list, the CH would notify it that its sampling rate for the next round must be reduced by $75\%$, since its level of correlation with sensor $6$ is 0.75.
Then the sensor $j$ (in this case $5$) is flagged as already notified. Thus, for the next sensor $j$ in the ordered list, if its matching sensor $i$ is already flagged. Instead of reducing its sampling rate proportionally to the level of correlation, it is reduced by (100 - $i$'s reduction $\%$). For instance, if the next sensor $j$ in the list is $3$, it matches with sensor $5$ in table \ref{table:corr1}, therefore, it's sampling rate will be reduced by $100-75=25\%$. And so on, until the last element in the ordered array.

\textbf{Algorithm 1 - line(50-56)} :However, some sensors may not appear in the second column of the table~\ref{table:corr1}, since they have not been matched with other sensors. Therefore, the CH looks for these sensor in the 1'st column of table~\ref{table:corr1}, and for each sensor $i$, it find their matching sensor $j$ in the second column, looks at how much the sampling rate was reduced for sensor $j$ and notifies sensor $i$ that its sampling rate must be reduced by (100 - sensor $j's$ reduction $\%$). 

The same explained operation is repeated at the end of each round. Therefore, enabling each sensor node to adjust its sampling rate according to its level of correlation with other sensors in the network. 
The algorithm 1 illustrates the proposed method that is implemented on the CH.

    \bigskip
	
    \hrule\vspace{5pt}
        	\noindent{\bfseries Algorithm 1}\quad STCSTA.\\ \vspace{-5pt}
        	\hrule\vspace{5pt}
        	\label{algo1}
        	\hspace*{\algorithmicindent} \textbf{Input:} $SRmax$ (1 sample/ X seconds) \\
        	\begin{algorithmic}[1]
        	\WHILE{$Energy \neq 0$}
        	\STATE $k \leftarrow 1$
        	\FOR {each sensor j in the cluster}
        	    \STATE receive the first value $v_{j}^0$ at the beginning of the round
            	\STATE $data[j][0] \leftarrow v_{j}^0$
            	\STATE $lastReceived[j] \leftarrow v_{j}^0 $
    	    \ENDFOR
        	
        	    \WHILE{! end of round}
        	        \IF{ nothing is received from sensor j after X seconds}
        	            \STATE $data[j][k] \leftarrow lastReceived[j]$
        	            
        	        \ELSIF {$v_{j}^n$ is received during the X seconds count}
        	            \STATE $data[j][k] \leftarrow v_{j}^n$
        	            \STATE $lastReceived[j] \leftarrow v_{j}^n$
    	            \ENDIF
        	        \STATE $k \leftarrow k+1$   
        	    \ENDWHILE
        	
        	    \IF {end of round}
        	        \FOR {i=1 to N}
        	            \FOR {j=i+1 to N}
        	            \STATE $corrArray[i][j] \leftarrow PearsonCorr(data[i][:],data[j][:])$
        	            \ENDFOR
        	        \ENDFOR
        	        
        	        \FOR {i=1 to N}
        	            \STATE $maxCorr[i][0] \leftarrow i $
        	            \STATE $[index,value] \leftarrow max(corrArray[i][:])$
        	            \STATE $maxCorr[i][1] \leftarrow index$
        	            \STATE $maxCorr[i][2] \leftarrow value;$
        	        \ENDFOR
        	        \STATE $k \leftarrow 1$
        	        
        	        \FOR {each element i $\in$ the second column of maxCorr}
        	            \IF {i $\notin$ first column of countOcc }
        	                \STATE $count \leftarrow$ count how many times i occures in the second column of maxCorr
        	                \STATE $countOcc[k][0] \leftarrow i$
        	                \STATE $countOcc[k][1] \leftarrow count$
        	                \STATE $k \leftarrow k+1$
        	            \ENDIF
        	        
        	        \ENDFOR
        	        
        	        \STATE order countOcc in ascending order according to the second column
        	        \STATE $k \leftarrow 1$
        	        \FOR {each element j $\in$ the first column of countOcc}
        	            \STATE $match \leftarrow maxCorr[j][1]$
        	            \IF {reduce[match-1] is empty}
        	                \STATE Notify sensor j that its sampling rate must be reduced by (maxcorr[j][2]*100)$\%$
        	                \STATE $reduce[j-1] \leftarrow (maxcorr[j][2]*100)$
        	            \ELSE    
        	                \STATE Notify sensor j that its sampling rate must be reduced by (100 - reduce[match-1])$\%$
        	                \STATE reduce[j-1]=100 - reduce[match-1]$\%$
        	            \ENDIF
        	        \ENDFOR
        	        \FOR {j=1 to N}
        	            \IF {reduce[j-1] is empty}
        	                \STATE $match \leftarrow maxCorr[j][1]$
        	                \STATE Notify sensor j that its sampling rate must be reduced by (100 - $reduce$[match-1])$\%$
    	               \ENDIF
        	        \ENDFOR
        	    
        	    \ENDIF
    	    \ENDWHILE
        	\end{algorithmic}
	\vspace{5pt}\hrule\vspace{10pt}


\subsection{Analysis Study}
\label{AS}

The objective of this algorithm is to create and manage a sampling rate balancing system based on the correlation degree between the nodes belonging to the same cluster. The idea is to match each sensor node with the one that correlates the most with, in such a way that, if one node of the paired couple reduces heavily its sampling rate, the other one keeps it high and vice versa, allowing them to compensate one another.
This compensation mechanism is crucial for the success of the reconstruction algorithm in term of minimizing the estimation error and increasing the quality of the replicated data. The latter relies on the correlation among sensor nodes in order to reconstruct the non-sampled measurements. Therefore, if highly correlated sensors are missing data simultaneously this would negatively affect the accuracy of the reconstructed measurement. When the balancing of non-sampled data is kept in check on the CH level, The reconstruction algorithm on the Sink will theoretically produce better estimations.

In this section, we will illustrate an example that explains our algorithm step by step. The latter provides a better analysis of what happens at the end of each round on the cluster head to better understand why and how this compensation system works.
Let us start by assuming that at the end of a given period, the CH has already computed the correlation between each pair of sensors belonging to the same cluster. In addition, we assume that the CH already matched each sensor with the one that correlates the most with and stored the results in a table similar to Table \ref{table:corr2}. The next step is to count for the sensors appearing in the second row of the table how many times it has been matched. For instance, sensor 7 has been matched 4 times, sensor 1 has been matched 2 times, and sensor 10, 9, 3, and 8 have been matched only one time.
The matched sensors are then ordered in ascending order according to how many times they have been matched.
the order will then be: $\{$sensor 8, sensor 3, sensor 9, sensor 10, sensor 1, sensor 7$\}$.

\def\arraystretch{1.5}%
\begin{table}[!htb]
\resizebox{\linewidth}{!}{
\begin{tabular}{|c|c|c|c|c|c|c|c|c|c|c|}
\hline
Sensor i                                                                         & 1  & 2  & 3  & 4  & 5  & 6  & 7  & 8  & 9  & 10 \\ \hline
\begin{tabular}[c]{@{}c@{}}Sensor j (has max \\ correlation with i)\end{tabular} & 8  & 1  & 7  & 3  & 9  & 1  & 10 & 7  & 7  & 7  \\ \hline
Correlation degree *$10^{-2}$                                                             & 78 & 69 & 54 & 92 & 85 & 72 & 79 & 83 & 89 & 90 \\ \hline
\end{tabular}
}
\caption{Table showing for each sensor its best match (maximum correlation) and the degree of correlation with this match}
\label{table:corr2}
\end{table}

Starting from the first sensor in the list (sensor 8) the CH looks for the sensor that it matches with. Looking at table \ref{table:corr2} we see that sensor 8 matches with sensor 7. The CH then checks whether the sampling rate of sensor 7 for the next round has been decided yet. If it is not the case the CH notes that the sensor 8 must reduce its sampling rate for the next round by 83\%, since the correlation degree for sensor 8 with its match is 0.83. The CH then follows the same procedure for the next sensor in the ordered list. sensor 3, 9, and 10 they all match with sensor 7 too, and since the sampling rate of sensor 7 has not been decided yet, their sampling rate will be reduced by 54\%, 89\%, and 90\% respectively for the next round.
Now the CH searches for the sensor that matches with the next sensor in the ordered list (sensor 1). Looking at table \ref{table:corr2} we see that it is sensor 8. However, the sampling rate of sensor 8 has been already decided to be reduced by 83\%, therefore instead of reducing the sampling rate of sensor 1 by 78\% it will be reduced by 100-83\%, therefore 17\% only. Same for sensor sensors 7, it matches with sensor 10, therefore its sampling rate must be reduced by 100-90\% (10\% only).

The next step is to adapt the sampling rate of the sensors that do not appear in the second row of the table, or in other words they have not been matched with other sensors in the cluster. in this example, the non-matched sensors are sensor 2,4,5 and 6.
Starting by sensor 2, its match is 1, therefore the sampling rate of sensor 2 for the next round must be reduced by 100-17\% (83\%), same for sensor 4,5, and 6 their sampling rate will be reduced respectively by 46\%, 11\%, and 83\%.

Before computing the percentage of the reduction in sampling rate, the matched sensors are first ordered in ascending order according to how many times they have been matched. The reason behind this crucial step can be explained as follows:
Let us suppose the list has not been ordered, and the CH started by sensor 7, which has been matched 4 times with 4 different sensors. The sampling rate of sensor 7 will be reduced by 79\%. Therefore, eventually, the sampling rates of sensors 3, 8, 9, and 10 will be reduced by 21\% only compared to 54\%, 83\%, 89\%, and 90\% respectively if the list was ordered. In consequence of not ordering the list first, the overall reduction in the sampling rate of the sensors would be reduced, which would lead to an increase in data transmission and energy consumption. Since sensor 7 can compensate for 4 other sensors, it is wise to leave it until the end, allowing the sensors that it matches with to reduce more their sampling rate.

A summary of the results is illustrated in table \ref{table:corr3}. We notice that if a sampling rate of a particular sensor is highly reduced, the one of the sensor that it correlates the most with will be proportionally and slightly reduced (e.g. sensors 2 and 1). This balanced reduction is meant to compensate for the matched sensor since the non-sampled values will eventually be derived mostly from its best match. Similarly, if the sampling rate of a sensor is slightly reduced, this will give more freedom to its match thus allowing it to highly reduce its sampling rate (e.g. sensors 5 and 9).

\def\arraystretch{1.5}%
\begin{table}[!htb]
\resizebox{\linewidth}{!}{
\begin{tabular}{|c|c|c|c|c|c|c|c|c|c|c|}
\hline
Sensor i                                                                        & 1  & 2  & 3  & 4  & 5  & 6  & 7  & 8  & 9  & 10 \\ \hline
SR reduction (\%)                                                               & 17 & 83 & 54 & 46 & 11 & 83 & 10 & 83 & 89 & 90 \\ \hline
\begin{tabular}[c]{@{}c@{}}Sensor j (has max\\ correlation with i)\end{tabular} & 8  & 1  & 7  & 3  & 9  & 1  & 10 & 7  & 7  & 7  \\ \hline
SR reduction (\%)                                                               & 83 & 17 & 10 & 54 & 89 & 17 & 90 & 10 & 10 & 10 \\ \hline
\end{tabular}
}
\caption{Table showing the \% of SR reduction for each sensor compared with its match}
\label{table:corr3}
\end{table}

\subsection{Reconstruction of the non sampled data}

In this section, the algorithm used to reconstruct non-sampled data is explained.
As mentioned earlier, the Sink detects and replaces non sampled data with a \enquote{Nan} value. After a certain period of time, let's say M rounds, defined by the user, the sink runs a reconstruction algorithm that can replace all the \enquote{Nan} values with estimations calculated using the spatial and temporal correlation among the data reported by the sensor nodes in the network. This algorithm it is deployed on the Sink instead of the CH due to its complexity. If deployed on CH it will consume a great amount of energy.
The reconstruction algorithm proposed in \cite{DynaMMo} essentially used to estimate missing data in co-evolving time series was adopted and adapted to suit our case.
Assuming after M rounds, the Sink would have stored in his Sink the following data-set:
\begin{center}
$SinkDataSet
=
\begin{bmatrix}
\centering
    x_{1}^{1} & x_{1}^{2} & x_{1}^{3} & \dots & x_{1}^{50} &  Nan  & x_{1}^{n*M} \\
    x_{2}^{1} & x_{2}^{2} & x_{2}^{3} & \dots & Nan & Nan & x_{2}^{n*M} \\
    \vdots & \vdots & \vdots & Nan & \vdots & \vdots & \vdots \\ 
    x_{N}^{1} & x_{N}^{2} & x_{N}^{3} & \dots & x_{1}^{50} & Nan & x_{N}^{n*M}
\end{bmatrix}
$

\begin{figure}[!htb]
    \centering
	\includegraphics[width=\linewidth]{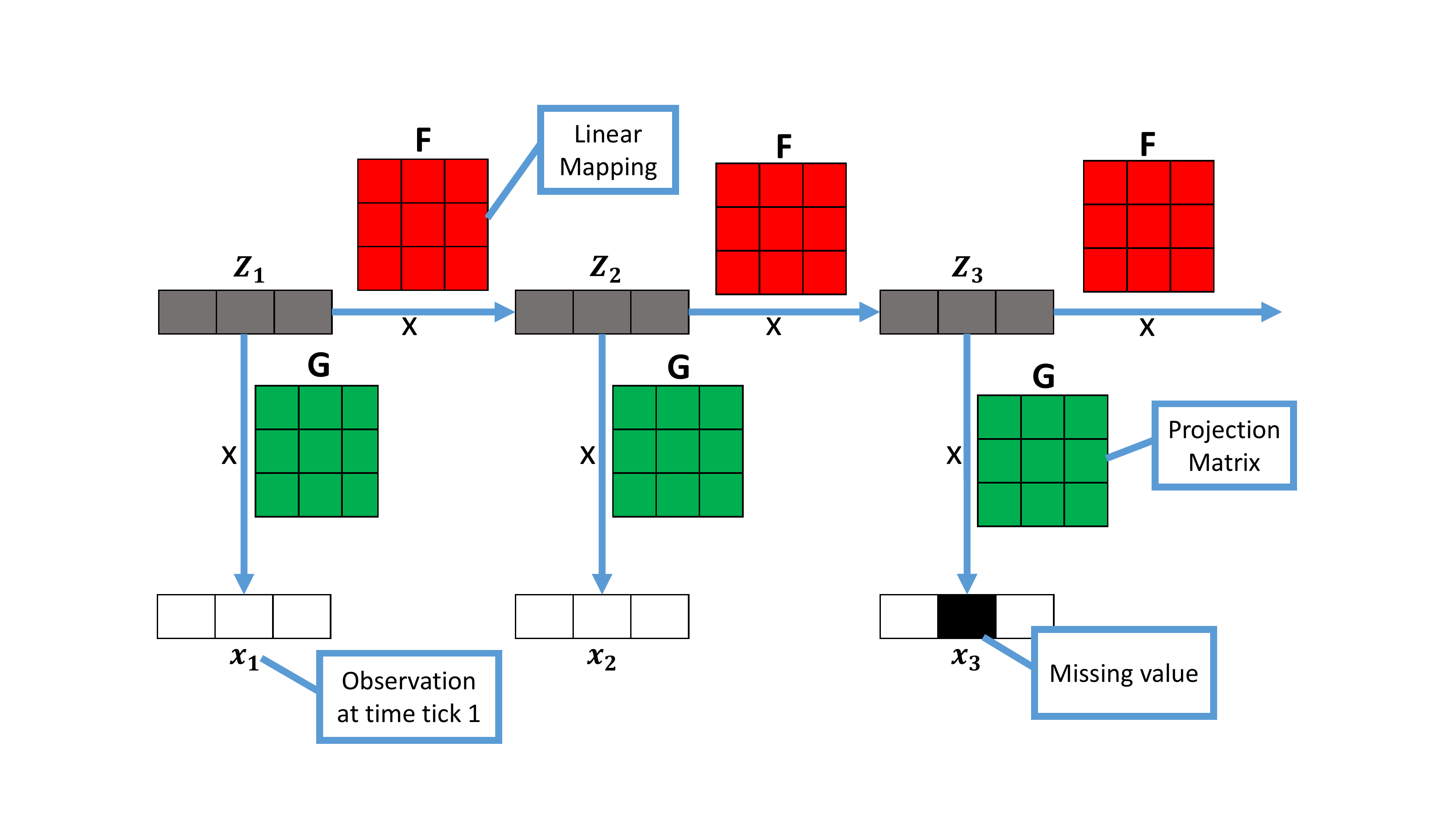}
	\caption{The probabilistic model}
	\label{fig:PM}
\end{figure}

\end{center}
A probabilistic model (Figure~\ref{fig:PM}) is built to estimate the expectation of missing values conditioned by the observed part. The model is built by initializing a latent variable $Z_1$, a linear mapping matrix $F$ and a projection matrix $G$, for the readers interested in how these values are initialized please refer to \cite{DynaMMo}. Afterward, using the linear mapping $F$ the algorithm can proceed to calculate the other $Z_n$ (n $\in$ [1,n*M]) by simply multiplying $Z_{n-1}*F$. Once all the values of $Z_n$ are calculated, the algorithm then estimates the observed and non-observed (Nan) values. This is achieved by multiplying each $Z_n$ by the projection Matrix $G$, which gives the predictions ($[x^{n}_1,...,x^{n}_N]$) of the values at the sampling time $n$.
Using the estimations, and the observed part, the algorithm then tries to maximize the log-likelihood of the observed sequences using an EM iterative algorithm \cite{EM} in order to update $F$ and $G$ and produce more accurate predictions. 
The same operation is repeated with the newly computed $F$ and $G$ until the number of iteration reaches a maximum value predefined by the user, or until the log-likelihood is no longer increasing.

\section{Experimental Results}
\label{ER}

We implemented our algorithm in addition to DPCAS~\cite{DPCAS} in a custom WSN simulator built in Matlab, and we conducted multiple experiments in order to evaluate and compare their performances. In the simulation, we used real sensor readings collected from a sensor network that was deployed at the Grand-St-Bernard pass between Switzerland and Italy~\cite{data}. The Network consisted of 23 sensors, each one of them collects 9 different environmental features with a fixed sampling rate of 1 sample every 2 minutes. We have chosen 4 out of these 9 features (ambient temperature $[C^{\circ}]$, Surface temperature $[C^{\circ}]$, relative humidity $[\%]$, and wind speed [m/s]), since the others are not complete. Environmental features are usually stationary, therefore, in addition to taking a sample every 2 minutes, and for a rigorous comparison, we set up two other scenarios, the first one, a sample is taken every 10 minutes instead, and the second one, a sample is taken every 20 minutes. In this way, the data will become \enquote{non-stationary} which makes it more realistic and harder for both algorithms to adapt to high variation in collected measurements.
The raw data set (sample every 2mins) consists of 10000 readings for each sensor, for the 1st scenario we will end up with 2000 readings instead, and 1000 readings for the second one.

In DPCAS the parameter $\epsilon$ defines the error tolerance of the application, the greater is $\epsilon$, the less is the amount of data that will be sampled and transmitted. However, the error of the estimated data will increase.
Therefore, the value of $\epsilon$ is the level of trade-off between the quality of the replicated data and the amount of sampled and transmitted measurements. In our experimentation, we set up five different values for $\epsilon$ ranging between $0.1$ and $0.5$ and we compare our approach to DPCAS for each value of $\epsilon$.

\subsection{Sampling and Transmission Reduction}
\label{STreduction}
In this section, we will explore and compare the effectiveness of each algorithm in reducing the number of both sampled and transmitted data in three different scenarios. As mentioned earlier, each sensor node collects 4 different environmental features (ambient temperature, surface temperature, relative humidity, and wind speed). For simplicity and better visualization of the results, all the figures will be illustrating the percentage of the aggregated sum of the data sampled and transmitted by the 23 nodes combined and for all features.

\begin{figure}[!htb]
  \caption{Average percentage of data sampled by each sensor node}
  \label{Fig:Sampled}
  \centering
    \includegraphics[width=\linewidth]{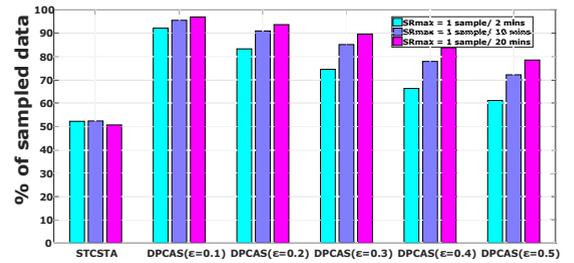}
\end{figure}

\begin{figure}[!htb]
  \caption{Average percentage of data transmitted by each sensor node}
  \label{Fig:Transmitted}
  \centering
    \includegraphics[width=\linewidth]{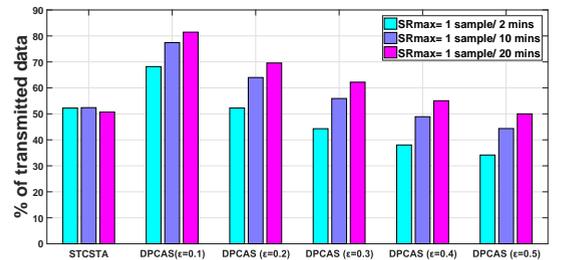}
\end{figure}

Figure~\ref{Fig:Sampled} and ~\ref{Fig:Transmitted} shows that on one hand, the bigger is the sampling interval between two consecutive measurements (higher variations in data), the greater is the average percentage of both sampled and transmitted data will be when DPCAS is deployed. On the other hand, when our approach (STCSTA) is deployed, the average percentage remains stable despite the level of variations in collected measurements, which makes it more robust, dynamic and tolerable to high variations. This is not the case for DPCAS however, its effectiveness can be significantly affected (a double-digit increase in sampled and transmitted data) depending on the type of data being collected. 
Moreover, Figure~\ref{Fig:Sampled} and~\ref{Fig:Transmitted} shows that STCSTA has the upper hand when it comes to reducing the number of both sampled and transmitted data.
For sampled data, Figure~\ref{Fig:Sampled} shows that STCSTA outperforms DPCAS in all scenarios and for all the values of $\epsilon$.
Figure~\ref{Fig:Transmitted} shows the average percentage of data transmitted by each one of the 23 nodes for both algorithms in 3 different scenarios and using different $\epsilon$ for DPCAS. The obtained results show the following: STCSTA outperforms DPCAS when $\epsilon \leq 0.2$ in all scenarios. However, for $\epsilon = 0.3$ DPCAS transmits less data in the first scenario ($SR_{max}$ = 1 sample/ 2mins), but more data in the other two scenarios ($SR_{max}$ = 1 sample/ 10 mins and 1 sample/ 20 mins). Finally, For $\epsilon = 0.4$ and $0.5$, DPCAS is slightly better in the first two scenarios. To sum it all up, the results in Figure \ref{Fig:Transmitted} show that STCSTA outperformed DPCAS 9 times, the latter outperformed STCSTA 5 times, and finally, we have 1 tie.

To conclude on this, when it comes to reducing the sampling and transmission rate, thus the energy consumed by the sampling activity $E_{sampling}$ and the transmission activity $E_{radio}$ STCSTA is more effective than DPCAS.

\subsection{Energy Consumption}
In this section, we present a comparison between the average energy consumed by the 23 sensor nodes when DPCAS and STCSTA are deployed. 

\begin{figure}[!htb]
  \caption{Average energy consumption of each sensor node}
  \label{EnergyConsumption}
  \centering
    \includegraphics[width=\linewidth]{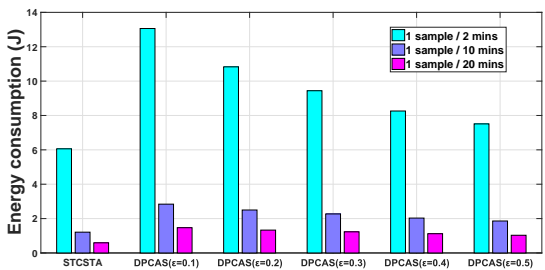}
\end{figure}

Previously, in section \ref{STreduction}, the obtained results clearly show that the $E_{radio}$ and $E_{sampling}$ are less when STCSTA is deployed since the amount of sampled and transmitted data is directly related to the energy consumed by the sampling and transmitting activities. However, according to Equation~\ref{eq:energyEq}, we still need to calculate $E_{logging}$ and $E_{Processing}$. This is where our approach shows a clear advantage. Knowing that in DPCAS an algorithm must be deployed on the node that handles 4 different sensors at a time. The node needs to perform reading and writing in the memory, and it needs to compute mathematical operation using the CPU. Therefore, the node will be consuming additional energy ($E_{logging}$ and $E_{Processing}$). However, for STCSTA, the node does not have to run an algorithm, nor to perform read and write in the memory, it simply collects a measurement using the integrated sensors, and directly transmits it to the CH. Therefore, no additional energy consumption is required.
Figure~\ref{EnergyConsumption} shows the average energy in Joule consumed by each one of the 23 deployed nodes. It is clear that our approach consumes approximately from $20\%$ up to $60\%$ less energy than DPCAS depending on the scenario and the value of $\epsilon$.

\subsection{Comparison with a baseline method}
The previously described results demonstrated that our approach STCSTA outperforms DPCAS in terms of energy preservation.
The DPCAS algorithm in \cite{DPCAS} was compared to two other approaches that use a similar technique, namely EDSAS \cite{EDSAS} and ASTCP \cite{ASTCP}. As mentioned in section \ref{RW}, the ASTCP algorithm was inspired by the EDSAS. Moreover, the DPCAS algorithm was inspired by both ASTCP and EDSAS. In this section, we will use the EDSAS as a baseline for comparison since it was the root algorithm that inspired both ASTCP and DPCAS. 
Table \ref{table:baseline} below shows the average energy consumed by each node in all scenarios and for the same value of $\epsilon$=0.1 used in \cite{DPCAS}. The obtained results are fairly similar to the ones obtained in \cite{DPCAS} and our approach remains better.

\def\arraystretch{1.5}%
\begin{table}[!htb]
\resizebox{\linewidth}{!}{
\begin{tabular}{|c|c|c|c|c|c|c|c|c|c|}
\hline
Algorithm                                                               & \multicolumn{3}{c|}{STCSTA} & \multicolumn{3}{c|}{DPCAS} & \multicolumn{3}{c|}{EDSAS} \\ \hline
\begin{tabular}[c]{@{}c@{}}Sampling Rate\\ 1 sample/ x min\end{tabular} & x=2     & x=10    & x=20    & x=2      & x=10   & x=20   & x=2      & x=10   & x=20   \\ \hline
Energy (J)                                                              & 6.06    & 1.21    & 0.59    & 13.06    & 2.84   & 1.47   & 13.38    & 2.92   & 1.52   \\ \hline
\end{tabular}
}
\caption{Table comparing STCSTA and DPCAS to the baseline EDSAS}
\label{table:baseline}
\end{table}

\subsection{The quality of the replicated data}
\label{QR}
\begin{figure}[!htb]
  \caption{Reconstructed wind speed signal}
  \label{fig:WspeedR}
  \centering
    \includegraphics[width=\linewidth]{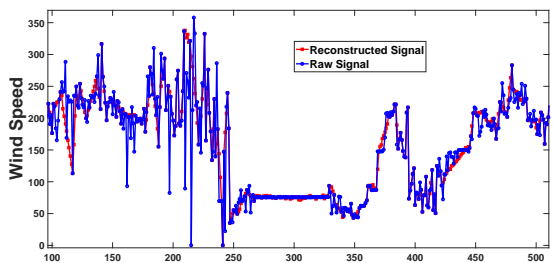}
\end{figure}

In order to measure the quality of the final set of data, we use the accuracy of the estimations as the validation criteria. Specifically, we use the Root Mean Square Error (RMSE) and the Mean Absolute Error (MAE) as an accuracy metric. Table~\ref{table:quality} shows the RMSE and MAE of the estimated data for the three scenarios.
For ambient temperature, surface temperature and relative humidity the errors are low. This is due to the fact that the spatial-temporal correlation of these features is strong, so the estimation algorithm can obtain an accurate and solid relationship based on mining correlation rules.
Table~\ref{table:quality} also shows that the error increases when the sampling interval widens. The bigger is the sampling interval, the weaker is the temporal correlation, therefore the harder is for the estimation algorithm to accurately estimate values. For Wind direction, the errors increase significantly but they are still proportionally low compared with the range of value for the wind speed (between 0 and 350 m/s). Wind speed has no spatial correlated with any other feature. Moreover, the wind speed value varies significantly between one sample and the other as shown in Figure \ref{fig:WspeedR}, therefore the temporal correlation is weak as well, that is why it has the highest error among other features.

\def\arraystretch{1.5}%
\begin{table}[!htb]
\centering
\caption {Quality of the reconstructed data} \label{tab:Error2000}
\label{table:quality}
\resizebox{\linewidth}{!}{
\begin{tabular}{cc|c|c|c|c|}
\cline{3-6}
&               & \textbf{Ambient Temp} & \textbf{Surface Temp} & \textbf{Relative Humidity} & \textbf{Wind Direction} \\ \hline

\multicolumn{1}{|c|}{\multirow{2}{*}{\textbf{\begin{tabular}[c]{@{}c@{}}1 sample/2 mins\end{tabular}}}} & \textbf{RMSE} & 1.12                  & 1.33                  & 2.7                       & 16.5                   \\ \cline{2-6} 
\multicolumn{1}{|c|}{}                                                                                              & \textbf{MAE}  & 0.71                  & 0.91                  & 1.89                       & 8.78                   \\ \hline
\multicolumn{1}{|c|}{\multirow{2}{*}{\textbf{\begin{tabular}[c]{@{}c@{}}1 sample/10 mins\end{tabular}}}} & \textbf{RMSE} & 1.26                  & 1.56                  & 3.68                       & 18.26                   \\ \cline{2-6} 
\multicolumn{1}{|c|}{}                                                                                              & \textbf{MAE}  & 0.74                  & 1.09                  & 2.55                       & 9.13                   \\ \hline
\multicolumn{1}{|c|}{\multirow{2}{*}{\textbf{\begin{tabular}[c]{@{}c@{}}1 sample/20 mins\end{tabular}}}} & \textbf{RMSE} & 1.43                  & 1.95                  & 4.78                       & 23.53                   \\ \cline{2-6} 
\multicolumn{1}{|c|}{}                                                                                              & \textbf{MAE}  & 0.87                  & 1.43                  & 3.21                       & 11.93                   \\ \hline

\end{tabular}
}
\end{table}

\begin{figure}[!htb]
  \caption{Reconstructed ambient temperature signal}
  \label{fig:AtempR}
  \centering
    \includegraphics[width=\linewidth]{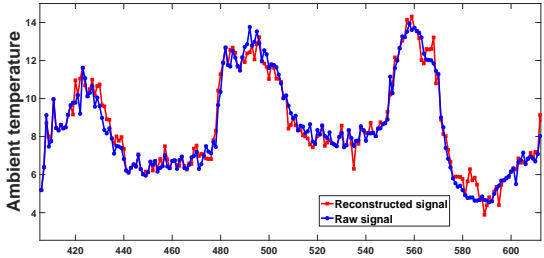}
\end{figure}

\begin{figure}[!htb]
  \caption{Reconstructed surface temperature signal}
  \label{fig:StempR}
  \centering
    \includegraphics[width=\linewidth]{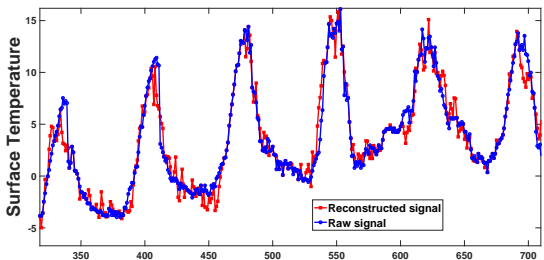}
\end{figure}

\begin{figure}[!htb]
  \caption{Reconstructed relative humidity signal}
  \label{fig:RhumR}
  \centering
    \includegraphics[width=\linewidth]{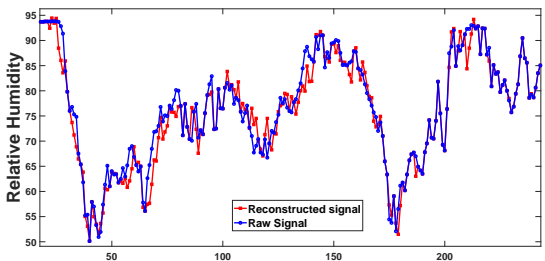}
\end{figure}

Figures \cref{fig:AtempR,fig:StempR,fig:RhumR,fig:WspeedR} shows a reconstructed signals for ambient temperature, surface temperature, relative humidity, and wind speed respectively. As shown in the figures, the data estimation (reconstruction) algorithm has been able to capture both the dynamics of the signal as well as the correlation across given inputs, therefore achieving a very satisfying reconstruction of the signals.
To conclude on the quality of the replicated data, simulation results presented in this section, demonstrated that the Sink is capable of reproducing the \enquote{non-sampled} data with a tolerable error margin. Thus, using our approach a sensor node can significantly reduce its sampling rate without affecting the integrity of the data.

\subsection{The Effect of The Sampling Strategy on Error Minimization}
The previous results have evaluated the efficiency of our proposed approach (STCSTA) in terms of reducing data transmission and energy consumption as well as the quality of the data replicated on the Sink. However, as previously explained in section \ref{AS}, the objective of our algorithm is to guarantee that the highly correlated sensors are not skipping data sampling simultaneously in order to reduce the reconstruction error. That was in theory, Therefore, in this section, we put the theory into practice in order to justify this claim.

Instead of building a list of matching sensors, ordering the list, and reducing the sampling rate of each sensor proportionally to its match. We eliminated the steps from line 30 and upward in Algorithm 1, only to allow a sensor to reduce its sampling rate according to its highest degree of correlation. For instance, let's assume that the sensor 1 has the highest correlation degree with sensor 5 (0.8). Without checking whether sensor 5 has already reduced its sampling rate or not, it will automatically reduce it by $80\%$. There is a chance that sensor 5 has already reduced its sampling rate lets say by $70\%$. Thus, both sensor 5 and 1 will skip sampling simultaneously which would, in theory, affect negatively the reconstruction algorithm, which will lead to an increase in the reconstruction error. We will be calling this method \enquote{The exaggerated sampling reduction} method. Table \ref{table:errortable2} shows the \% of increase in the reconstruction error when this method is applied. We notice that the Reconstruction error increases significantly in all scenarios and for all environmental features, which justifies our controlled sampling strategy.

\def\arraystretch{1.5}%
\begin{table}[!htb]
\centering
\caption {Percentage of increase in reconstruction error (the exaggerated sampling reduction method)}
\label{table:errortable2}
\resizebox{\linewidth}{!}{
\begin{tabular}{cc|c|c|c|c|}
\cline{3-6}
                                                        &      & Ambient Temp & Surface Temp & Relative Humidity & Wind Direction \\ \hline
\multicolumn{1}{|c|}{\multirow{2}{*}{1 sample/2 mins}}  & RMSE & 16.9 \%      & 34.5 \%      & 20 \%             & 45.3 \%        \\ \cline{2-6} 
\multicolumn{1}{|c|}{}                                  & MAE  & 12.6 \%      & 41.7 \%      & 16.93 \%          & 23.4 \%        \\ \hline
\multicolumn{1}{|c|}{\multirow{2}{*}{1 sample/10 mins}} & RMSE & 26.9 \%      & 44.8 \%      & 35.8 \%           & 52.7 \%        \\ \cline{2-6} 
\multicolumn{1}{|c|}{}                                  & MAE  & 39.1 \%      & 52.3 \%      & 29.4 \%           & 75.2 \%        \\ \hline
\multicolumn{1}{|c|}{\multirow{2}{*}{1 sample/20 mins}} & RMSE & 25.8 \%      & 48.2 \%      & 50.2 \%           & 36.0 \%        \\ \cline{2-6} 
\multicolumn{1}{|c|}{}                                  & MAE  & 25.2 \%      & 44.0 \%      & 45.7 \%           & 59.2 \%        \\ \hline
\end{tabular}
}
\end{table}

\subsection{Scalability and Limitations }

Obviously, the scalability of such a network depends on the computational power of the CH and its memory capacity. A more powerful CPU and big memory size mean that the CH could handle a large number of sensors simultaneously. The weaker is the CPU and the smaller is the memory size, the fewer nodes a CH can handle. A great number of devices that can be used as a CH are currently available in the market, they all have different features and characteristics. One can find cheap less powerful CH device for personal use or an expensive and powerful device for commercial use. Therefore, the choice of the CH depends on the size of the network a user wants to deploy. A network consisting of thousands of nodes will certainly need a powerful CH. However, a network consisting of a few hundred or tens of nodes could work just fine with a less powerful CH.

\begin{figure}[!htb]
  \caption{Memory size needed for the first part of the Algorithm (line 1-28)}
  \label{Fig:116}
  \centering
    \includegraphics[width=\linewidth]{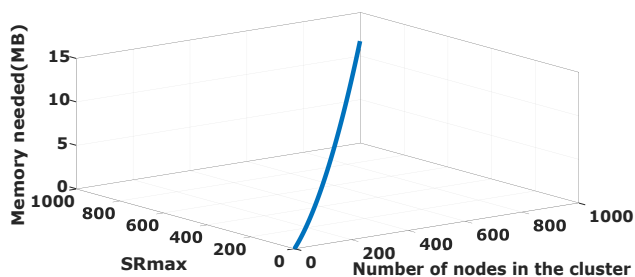}
\end{figure}

\begin{figure}[!htb]
  \caption{Memory size needed for the second part of the Algorithm (line 23-57)}
  \label{Fig:17up}
  \centering
    \includegraphics[width=\linewidth]{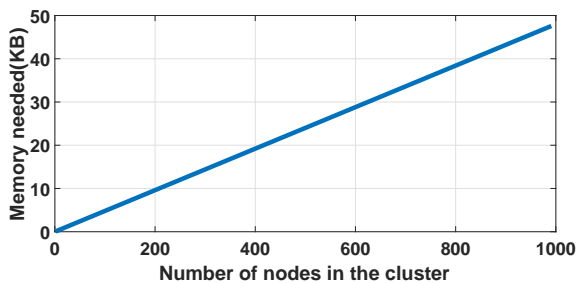}
\end{figure}

Our proposed algorithm is not very complex though, it has a complexity that is linear in time (O(n)). This linear complexity allows the CH to handle a large number of nodes with minimal computational power. Regarding the memory size required by the STCSTA, assuming that the number of nodes in the cluster is $N$, and each value is encoded into 8bytes.
\begin{itemize}
    \item $8 \times (N(SR_{max}+\frac{1}{2}N + 4)+1)$ bytes is the memory size required by the Algorithm1 from line 1-28. Figure~\ref{Fig:116} shows the memory size needed by the CH in function of $SR_{max}$ and the number of nodes belonging to the cluster.
    \item $8 (\times 6N+1$) bytes is the memory size required by the Algorithm1 from line 23-57 if we assume that the matching sensors are at maximum equal to the number of sensors in the cluster.  Figure~\ref{Fig:17up} shows the maximum memory size needed by the CH in function of the number of nodes belonging to the cluster.
\end{itemize} 
The maximum memory size required by the CH is $8 \times Max(N(SR_{max}+\frac{1}{2}N + 4)+1 , 6N+1)$ bytes, since the values stored in the first part of the Algorithm (1-17), could be cleared once the sensors have been matched (Algorithm 1, line 17-28).

Nevertheless, the greater is the number of nodes belonging to the same cluster, the better is the correlation among these nodes, the fewer data a sensor will sample and transmit which eventually leads to less energy consumption. Therefore, the number of sensors belonging to the same cluster should be maximized in function of its computational and memory resources.

As for the limitation of our proposed algorithm, it is evident when there is no or little correlation among the collected measurements, the sampling rate of the sensors will be always kept high. Since the role of this algorithm is to minimize the sampling rate of the sensor node, it will not be as efficient as it should be. 

\section{Conclusion}
We proposed in this paper a sampling and transmission rate adaptation algorithm for cluster-based sensor networks. This algorithm is deployed on the Cluster-Head (CH) and it operates in rounds. The latter controls the sampling rate of each individual sensor node by increasing it or decreasing it according to its spatial correlation with other sensors in the network. Moreover, we adopted and adapted a data reconstruction algorithm that is implemented on the Sink station. The latter can identify the \enquote{non-sampled} data that are not collected due to a decrease in the sampling rate of a specific sensor and it estimates them using an EM iterative approach that is capable of capturing the temporal and spatial correlation among the reported measurements.
We presented experimentation that we have conducted on real sensor data of a network that was deployed at the Grand-St-Bernard pass located between Switzerland and Italy. We have compared our approach with a recent data reduction technique that combines both adaptive sampling and transmission reduction. The obtained results demonstrate that our proposal is better at reducing the energy consumption of the sensor node, thus extending the operational lifetime of the network while preserving the integrity and the quality of the data.

For future work, we aim to tune better the Algorithm deployed on the CH by incorporating other attributes to determine the optimal sampling rate of each individual sensor. Moreover, we will explore the possibility of adding a compression phase between the CH and the workstation in order to reduce more the amount of transmitted data.

\section{Acknowledgement}
This work is partially funded by the EIPHI Graduate School (contract "ANR-17-EURE-0002"), the France-Suisse Interreg RESponSE project, 
 Lebanese University Research Program (Number: 4/6132), and EPSRC PETRAS 2 (EP/S035362/1).

\bibliographystyle{IEEEtran}
\bibliography{References}

\begin{thebibliography}{10}
\providecommand{\url}[1]{#1}
\csname url@samestyle\endcsname
\providecommand{\newblock}{\relax}
\providecommand{\bibinfo}[2]{#2}
\providecommand{\BIBentrySTDinterwordspacing}{\spaceskip=0pt\relax}
\providecommand{\BIBentryALTinterwordstretchfactor}{4}
\providecommand{\BIBentryALTinterwordspacing}{\spaceskip=\fontdimen2\font plus
\BIBentryALTinterwordstretchfactor\fontdimen3\font minus
  \fontdimen4\font\relax}
\providecommand{\BIBforeignlanguage}[2]{{%
\expandafter\ifx\csname l@#1\endcsname\relax
\typeout{** WARNING: IEEEtran.bst: No hyphenation pattern has been}%
\typeout{** loaded for the language `#1'. Using the pattern for}%
\typeout{** the default language instead.}%
\else
\language=\csname l@#1\endcsname
\fi
#2}}
\providecommand{\BIBdecl}{\relax}
\BIBdecl

\bibitem{eSampling}
M.~Z.~A. Bhuiyan, J.~Wu, G.~Wang, T.~Wang, and M.~M. Hassan, ``e-sampling:
  Event-sensitive autonomous adaptive sensing and low-cost monitoring in
  networked sensing systems,'' \emph{ACM Trans. Auton. Adapt. Syst.}, vol.~12,
  no.~1, pp. 1:1--1:29, Mar. 2017.

\bibitem{Asampling2}
A.~Masoum, N.~Meratnia, and P.~J.~M. Havinga, ``A decentralized quality aware
  adaptive sampling strategy in wireless sensor networks,'' in \emph{2012 9th
  International Conference on Ubiquitous Intelligence and Computing and 9th
  International Conference on Autonomic and Trusted Computing}, Sept 2012, pp.
  298--305.

\bibitem{TIIMakhoul}
H.~Harb and A.~Makhoul, ``Energy efficient sensor data collection approach for
  industrial process monitoring,'' \emph{IEEE Transactions on Industrial
  Informatics}, no.~2, pp. 661 -- 672, feb 2018.

\bibitem{AdhocMak}
A.~Makhoul, H.~Moustafa~Harb, and D.~Laiymani, ``Residual energy-based adaptive
  data collection approach for periodic sensor networks,'' \emph{Ad Hoc
  Networks}, vol.~35, pp. 149--160, dec 2015.

\bibitem{ASTR}
G.~B. Tayeh, A.~Makhoul, D.~Laiymani, and J.~Demerjian, ``A distributed
  real-time data prediction and adaptive sensing approach for wireless sensor
  networks,'' \emph{Pervasive and Mobile Computing}, vol.~49, pp. 62 -- 75,
  2018.

\bibitem{HLMS}
L.~Tan and M.~Wu, ``Data reduction in wireless sensor networks: A hierarchical
  lms prediction approach,'' \emph{IEEE Sensors Journal}, vol.~16, no.~6, pp.
  1708--1715, March 2016.

\bibitem{LMS}
S.~Santini and K.~R{\"o}mer, ``An adaptive strategy for quality-based data
  reduction in wireless sensor networks,'' in \emph{Proceedings of the 3rd
  International Conference on Networked Sensing Systems}, 2006, pp. 29--36.

\bibitem{OSSLMS}
M.~Wu, L.~Tan, and N.~Xiong, ``Data prediction, compression, and recovery in
  clustered wireless sensor networks for environmental monitoring
  applications,'' \emph{Information Sciences}, vol. 329, no. Supplement C, pp.
  800 -- 818, 2016.

\bibitem{TR}
G.~B. Tayeh, A.~Makhoul, J.~Demerjian, and D.~Laiymani, ``A new autonomous data
  transmission reduction method for wireless sensors networks,'' in \emph{2018
  IEEE Middle East and North Africa Communications Conference (MENACOMM)},
  April 2018, pp. 1--6.

\bibitem{CP1}
\BIBentryALTinterwordspacing
B.~Alturki, S.~Reiff-Marganiec, and C.~Perera, ``A hybrid approach for data
  analytics for internet of things,'' in \emph{Proceedings of the Seventh
  International Conference on the Internet of Things}, ser. IoT '17.\hskip 1em
  plus 0.5em minus 0.4em\relax New York, NY, USA: ACM, 2017, pp. 7:1--7:8.
  [Online]. Available: \url{http://doi.acm.org/10.1145/3131542.3131558}
\BIBentrySTDinterwordspacing

\bibitem{DPCAS}
L.~C. Monteiro, F.~C. Delicato, L.~Pirmez, P.~F. Pires, and C.~Miceli, ``Dpcas:
  Data prediction with cubic adaptive sampling for wireless sensor networks,''
  in \emph{International Conference on Green, Pervasive, and Cloud
  Computing}.\hskip 1em plus 0.5em minus 0.4em\relax Springer, 2017, pp.
  353--368.

\bibitem{Compression1}
M.~T. Nguyen, K.~A. Teague, and N.~Rahnavard, ``Ccs: Energy-efficient data
  collection in clustered wireless sensor networks utilizing block-wise
  compressive sensing,'' \emph{Computer Networks}, vol. 106, pp. 171 -- 185,
  2016.

\bibitem{compression3}
X.~Liu, Y.~Zhu, L.~Kong, C.~Liu, Y.~Gu, A.~V. Vasilakos, and M.~Wu, ``Cdc:
  Compressive data collection for wireless sensor networks,'' \emph{IEEE
  Transactions on Parallel and Distributed Systems}, vol.~26, no.~8, pp.
  2188--2197, Aug 2015.

\bibitem{JA1}
J.~Azar, R.~Darazi, C.~Habib, A.~Makhoul, and J.~Demerjian, ``Using dwt lifting
  scheme for lossless data compression in wireless body sensor networks,'' in
  \emph{2018 14th International Wireless Communications Mobile Computing
  Conference (IWCMC)}, June 2018, pp. 1465--1470.

\bibitem{JA2}
J.~Azar, A.~Makhoul, R.~Darazi, J.~Demerjian, and R.~Couturier, ``On the
  performance of resource-aware compression techniques for vital signs data in
  wireless body sensor networks,'' in \emph{2018 IEEE Middle East and North
  Africa Communications Conference (MENACOMM)}, April 2018, pp. 1--6.

\bibitem{Agg1}
J.~Yang, S.~Tilak, and T.~S. Rosing, ``An interactive context-aware power
  management technique for optimizing sensor network lifetime,'' in
  \emph{SENSORNETS}, 2016, pp. 69--76.

\bibitem{Agg1Mak}
H.~Harb, A.~Makhoul, R.~Couturier, and S.~Tawbi, ``Comparison of different data
  aggregation techniques in distributed sensor networks,'' \emph{IEEE Access},
  vol.~5, pp. 4250 -- 4263, mar 2017.

\bibitem{Agg2Mak}
H.~Harb, A.~Makhoul, D.~Laiymani, and A.~Jaber, ``A distance-based data
  aggregation technique for periodic sensor networks,'' \emph{ACM Transactions
  on Sensor Networks}, vol.~13, no.~4, p. 32 (40 pages), sep 2017.

\bibitem{CP2}
D.~{Kaur}, G.~S. {Aujla}, N.~{Kumar}, A.~Y. {Zomaya}, C.~{Perera}, and
  R.~{Ranjan}, ``Tensor-based big data management scheme for dimensionality
  reduction problem in smart grid systems: Sdn perspective,'' \emph{IEEE
  Transactions on Knowledge and Data Engineering}, vol.~30, no.~10, pp.
  1985--1998, Oct 2018.

\bibitem{Agg2}
A.~Makhoul and H.~Harb, ``Data reduction in sensor networks: Performance
  evaluation in a real environment,'' \emph{IEEE Embedded Systems Letters},
  vol.~9, no.~4, pp. 101--104, 2017.

\bibitem{EnergyModel}
M.~N. Halgamuge, M.~Zukerman, K.~Ramamohanarao, and H.~L. Vu, ``An estimation
  of sensor energy consumption,'' \emph{Progress in Electromagnetics Research},
  vol.~12, pp. 259--295, 2009.

\bibitem{En2}
H.-Y. Zhou, D.-Y. Luo, Y.~Gao, and D.-C. Zuo, ``Modeling of node energy
  consumption for wireless sensor networks,'' \emph{Wireless Sensor Network},
  vol.~3, no.~01, p.~18, 2011.

\bibitem{En3}
W.~Du, F.~Mieyeville, and D.~Navarro, ``Modeling energy consumption of wireless
  sensor networks by systemc,'' in \emph{Systems and Networks Communications
  (ICSNC), 2010 Fifth International Conference on}.\hskip 1em plus 0.5em minus
  0.4em\relax IEEE, 2010, pp. 94--98.

\bibitem{RLS}
Q.~A. Bakhtiar, K.~Makki, and N.~Pissinou, ``Data reduction in low powered
  wireless sensor networks,'' in \emph{Wireless Sensor Networks-Technology and
  Applications}.\hskip 1em plus 0.5em minus 0.4em\relax InTech, 2012.

\bibitem{DBP}
U.~Raza, A.~Camerra, A.~L. Murphy, T.~Palpanas, and G.~P. Picco, ``Practical
  data prediction for real-world wireless sensor networks,'' \emph{IEEE
  Transactions on Knowledge and Data Engineering}, vol.~27, no.~8, pp.
  2231--2244, Aug 2015.

\bibitem{ASTCP}
N.~Al-Hoqani and S.-H. Yang, ``Adaptive sampling for wireless household water
  consumption monitoring,'' \emph{Procedia Engineering}, vol. 119, pp.
  1356--1365, 2015.

\bibitem{EDSAS}
M.~Gupta, L.~V. Shum, E.~Bodanese, and S.~Hailes, ``Design and evaluation of an
  adaptive sampling strategy for a wireless air pollution sensor network,'' in
  \emph{Local Computer Networks (LCN), 2011 IEEE 36th Conference on}.\hskip 1em
  plus 0.5em minus 0.4em\relax IEEE, 2011, pp. 1003--1010.

\bibitem{ST1}
L.~A. Villas, A.~Boukerche, D.~L. Guidoni, H.~A. De~Oliveira, R.~B. De~Araujo,
  and A.~A. Loureiro, ``An energy-aware spatio-temporal correlation mechanism
  to perform efficient data collection in wireless sensor networks,''
  \emph{Computer Communications}, vol.~36, no.~9, pp. 1054--1066, 2013.

\bibitem{ST2}
K.~Karuppasamy and V.~Gunaraj, ``Optimizing sensing quality with coverage and
  lifetime in wireless sensor networks,'' \emph{International Journal of
  Engineering Research and Technology}, vol.~2, no.~2, pp. 1--7, 2013.

\bibitem{ST3}
S.~Dhimal and K.~Sharma, ``Energy conservation in wireless sensor networks by
  exploiting inter-node data similarity metrics,'' \emph{International Journal
  of Energy, Information and Communications}, vol.~6, no.~2, pp. 23--32, 2015.

\bibitem{ST4}
H.~Harb and A.~Makhoul, ``Energy-efficient scheduling strategies for minimizing
  big data collection in cluster-based sensor networks,'' \emph{Peer-to-Peer
  Networking and Applications}, pp. 1--15, 2018.

\bibitem{ST5}
H.~Harb, A.~Makhoul, A.~Jaber, and S.~Tawbi, ``Energy efficient data collection
  in periodic sensor networks using spatio-temporal node correlation,''
  \emph{International Journal of Sensor Networks (to appear)}, 2019.

\bibitem{CP3}
C.~{Perera}, D.~S. {Talagala}, C.~H. {Liu}, and J.~C. {Estrella},
  ``Energy-efficient location and activity-aware on-demand mobile distributed
  sensing platform for sensing as a service in iot clouds,'' \emph{IEEE
  Transactions on Computational Social Systems}, vol.~2, no.~4, pp. 171--181,
  Dec 2015.

\bibitem{CP4}
P.~P. {Jayaraman}, C.~{Perera}, D.~{Georgakopoulos}, and A.~{Zaslavsky},
  ``Efficient opportunistic sensing using mobile collaborative platform
  mosden,'' in \emph{9th IEEE International Conference on Collaborative
  Computing: Networking, Applications and Worksharing}, Oct 2013, pp. 77--86.

\bibitem{DynaMMo}
L.~Li, J.~McCann, N.~S. Pollard, and C.~Faloutsos, ``Dynammo: Mining and
  summarization of coevolving sequences with missing values,'' in
  \emph{Proceedings of the 15th ACM SIGKDD international conference on
  Knowledge discovery and data mining}.\hskip 1em plus 0.5em minus 0.4em\relax
  ACM, 2009, pp. 507--516.

\bibitem{EM}
Z.~Ghahramani and M.~I. Jordan, ``Supervised learning from incomplete data via
  an em approach,'' in \emph{Advances in neural information processing
  systems}, 1994, pp. 120--127.

\bibitem{data}
{Sensorscope}, \url{https://lcav.epfl.ch/page-145180-en.html}, 2007, online;
  accessed 13 Septembre 2018.

\end{thebibliography}

\vskip 0pt plus -1fil
\begin{IEEEbiography}
    [{\includegraphics[width=1in,height=1.25in,clip,keepaspectratio]{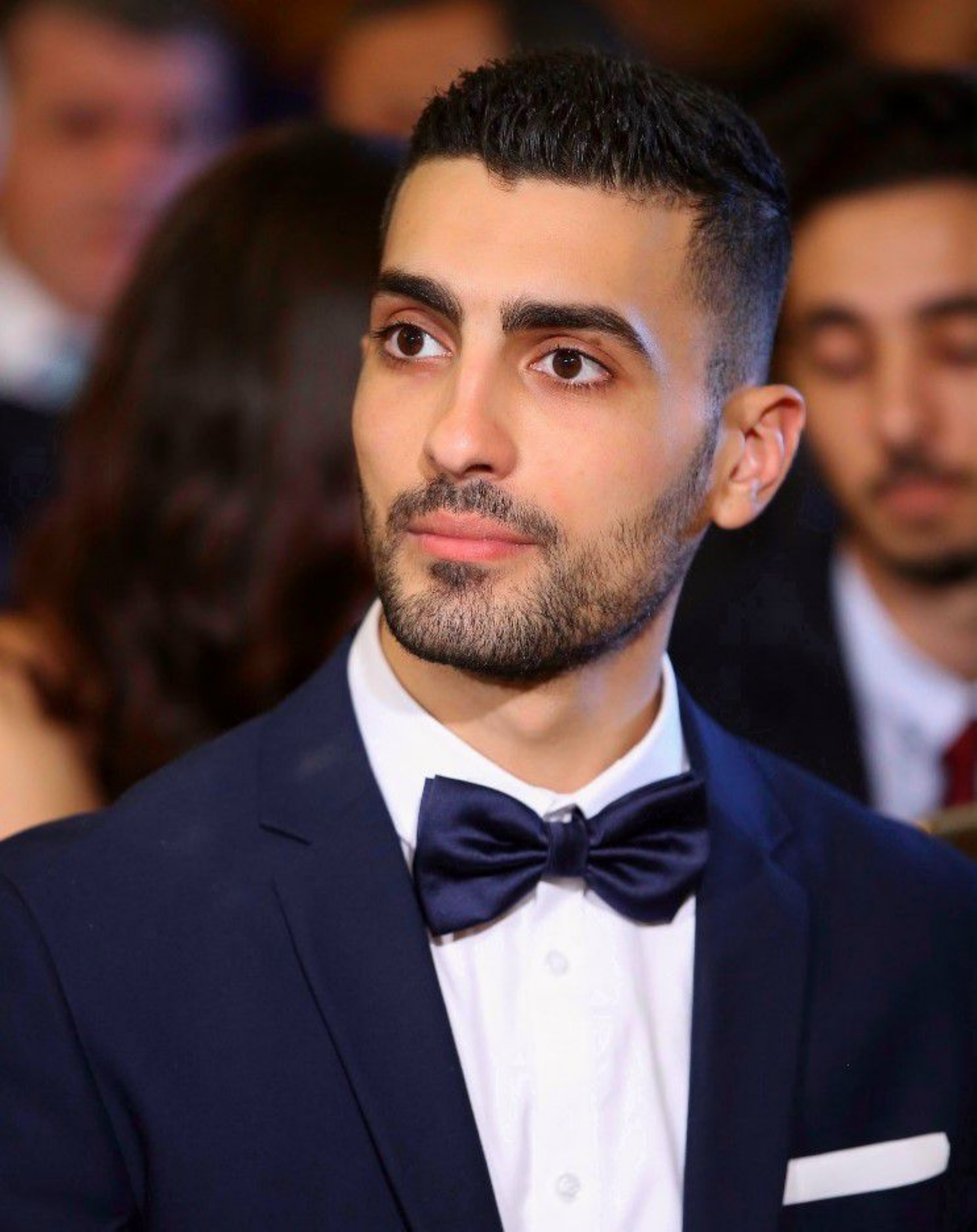}}]{Gaby BOU TAYEH}
    Received an M.S degree in mobile and distributed computing from the University of Franche-Comt\'{e} (UFC), Belfort, France in 2017. He is currently a Ph.D. Student at the University of Franche-Comt\'{e}. 
    His research interests include Wireless Sensor Networks, LoRaWan, Data Science, and Big Data.
\end{IEEEbiography}
\vskip 0pt plus -1fil
\begin{IEEEbiography}
    [{\includegraphics[width=1in,height=1.25in,clip,keepaspectratio]{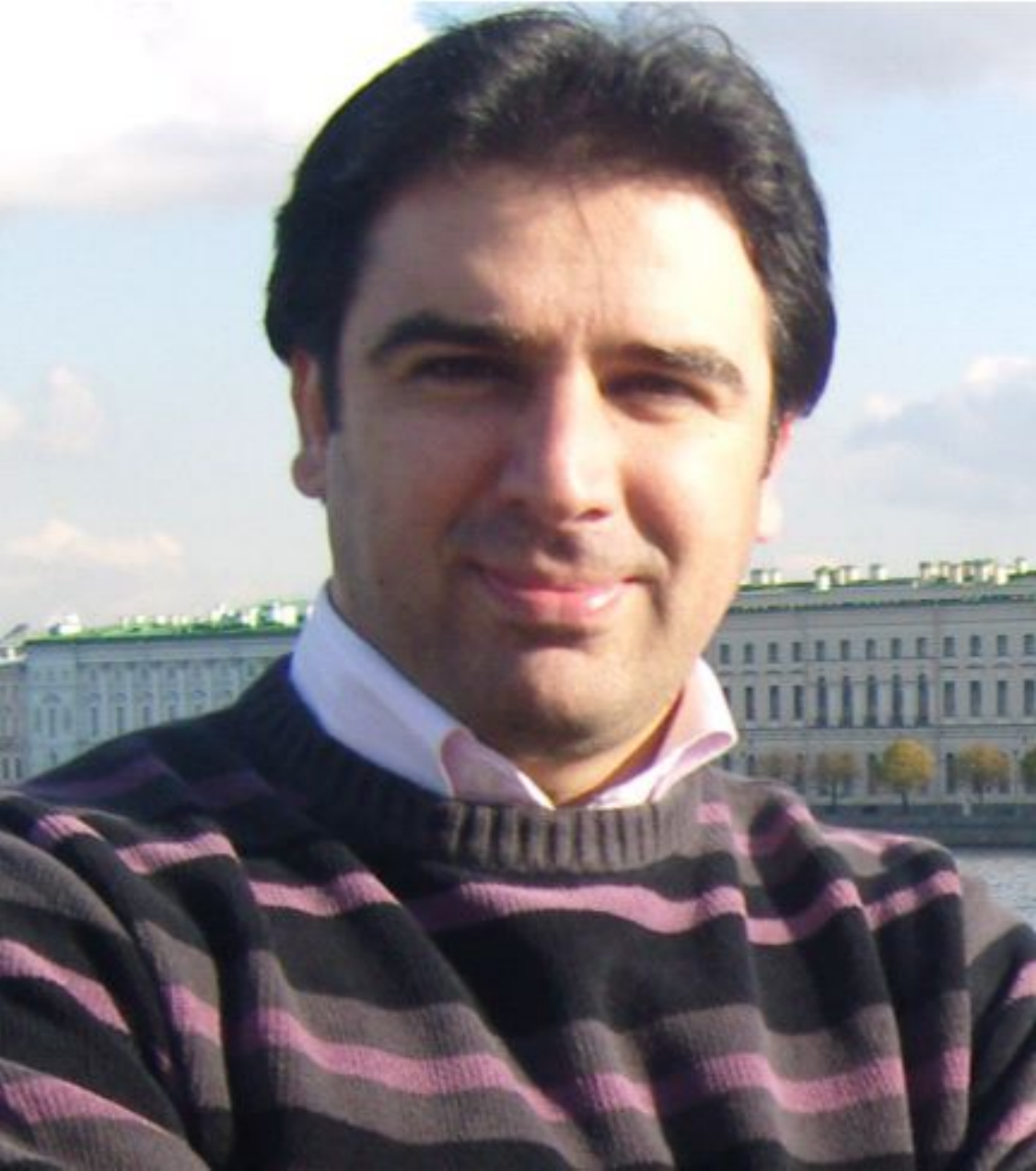}}]{Abdallah MAKHOUL}
    Received the M.S. degree in computer science from INSA Lyon, Lyon, France, in 2005, and the Ph.D. degree from the University of Franche-Comté, Belfort, France, in 2008. Since 2009, he has been an Associate Professor with the University of Franche-Comté. His research interests include Internet of Things, structural health monitoring, and real-time issues in Wireless Sensor Networks. Dr. MAKHOUL has been a TPC chair and member of several networking conferences and workshops and a Reviewer for several international journals.
\end{IEEEbiography}
\vskip 0pt plus -1fil
\begin{IEEEbiography}
    [{\includegraphics[width=1in,height=1.25in,clip,keepaspectratio]{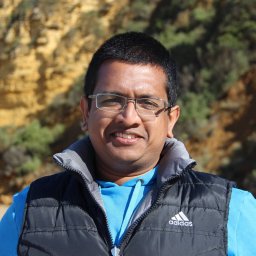}}]{Charith Perera} is a Lecturer (Assistant Professor) at Cardiff University, UK. He received the B.Sc. degree (Hons.) in computer science from Staffordshire University, U.K., the MBA degree in business administration from the University of Wales, Cardiff, U.K., and the Ph.D. degree in computer science from The Australian National University, Canberra, Australia. He was with the Information Engineering Laboratory, ICT Centre, CSIRO. He is currently a Lecturer (Assistant Professor) with Cardiff University, U.K. His research interests are Internet of Things, sensing as a service, privacy, middleware platforms, and sensing infrastructure. He is a member of ACM.

\end{IEEEbiography}
\vskip 0pt plus -1fil
\begin{IEEEbiography}
    [{\includegraphics[width=1in,height=1.25in,clip,keepaspectratio]{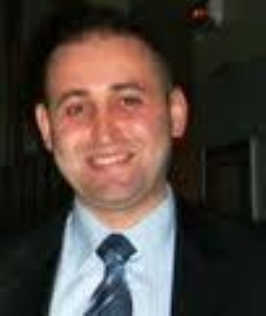}}]{Jacques DEMERJIAN}
     has received his Ph.D. degree in Network and Computer Science from TELECOM ParisTech (ENST-Paris) in 2004. Dr. DEMERJIAN is an Associate Professor at the Faculty of Sciences at the Lebanese University in Lebanon. His main research interests include Human-Computer Interaction, Streaming Data Quality and Summarization, Mobile Cloud Computing, VANET, Body Sensor Network, Data Mining and Wired and Wireless Network Security.
\end{IEEEbiography}

\EOD

\end{document}